\documentclass[amsmath,amssymb,showpacs,twocolumn]{revtex4}
\usepackage{mathrsfs}
\usepackage{graphicx}  
\usepackage{cases}

\begin{document}

\title{A variational surface hopping algorithm for the sub-Ohmic spin-boson model}

\author{Yao Yao$^{1,2}$ and Yang Zhao$^1$\footnote{Electronic address:~\url{YZhao@ntu.edu.sg}}}

\address{
$^1$Division of Materials Science, Nanyang Technological University, 50 Nanyang Avenue, Singapore 639798\\
$^2$State Key Laboratory of Surface Physics and Department
of Physics, Fudan University, Shanghai 200433, China
}

\date{\today}
\begin{abstract}
The Davydov D$_1$ ansatz, which assigns individual bosonic trajectories
to each spin state, is an efficient, yet extremely accurate trial state for time-dependent
variation of the sub-Ohmic spin-boson model [J.~Chem.~Phys.~{\bf 138}, 084111 (2013)]. A surface hopping algorithm is
developed employing the Davydov
D$_1$ ansatz to study the spin dynamics with a sub-Ohmic
bosonic bath.
The algorithm takes into account both coherent and incoherent dynamics
of the population evolution in a
unified manner, and
compared with
semiclassical surface hopping algorithms, hopping rates calculated in this work follow more closely
the Marcus formula.
\end{abstract}


\maketitle

\section{introduction}

Rapid advances of time-resolved two-dimensional (2D) optical
spectroscopy in the recent years have greatly stimulated
theoretical interest on the long-lived quantum coherence in
light-harvesting systems \cite{PS1,PS2,PS3} such as
the Fenna-Matthews-Olson (FMO) complex in green sulfur bacteria \cite{FMO1,FMO2}.
Information provided by 2D spectroscopy
on excited state populations and coherences allows greater
latitude for studying the physics of
decoherence and dissipation. It was claimed recently that the quantum coherence
lasts for more than 400 fs at room temperature during the
excitation transfer process \cite{PS3},
a surprising finding as the length of the decoherence time was
estimated to be much shorter.
Subsequently, various
system-bath models have been reexamined,
among which much attention has been attracted by the spin-boson model (SBM)\cite{SBM1,SBM2} with
the spin denoting an effective one-exciton state residing in a
pair of interacting chromophores \cite{JPCM,JPCL}.
The SBM Hamiltonian can be written as
\begin{eqnarray}
\hat{H}=\frac{\epsilon}{2}\sigma^z-\frac{\Delta}{2}\sigma^x+\sum_l\omega_l
b^{\dag}_lb_l+\frac{\sigma^z}{2}\sum_l\lambda_l(b^{\dag}_l+b_l),\label{hami}
\end{eqnarray}
where $\hbar$ is set to unity, $\sigma^z$ and $\sigma^x$ are the
usual Pauli operators, $\epsilon$ is the spin bias, $\Delta$ is the
tunneling constant, $\omega_l$ is the frequency of the $l$-th boson
mode, $b^{\dag}_l (b_l)$ is the boson creation (annihilation) operator of
the $l$-th mode, and $\lambda_l$ labels the coupling strength of the spin to
the $l$-th mode. The frequency cutoff of bosons is $\omega_c$, and
the spectral function is given as
$J(\omega)=2\pi\alpha\omega^{1-s}_c\omega^s$. The Ohmic bosonic bath
is specified by s=1, while $s<1$ denotes the sub-Ohmic bath. The
coherent-incoherent transition of the spin boson model for both
Ohmic and sub-Ohmic bath has been extensively studied
\cite{ChinPRB,Kast,WangNew,Zhao3,PRB94,PRA2,PRB3,JCP,JPCL,NJP,PRA,PRB2,wuning}. The sub-Ohmic
regime is especially interesting as sustained quantum coherence is
known to exist in this regime. For example, by the numerically exact quasiadiabatic propagator path
integral method (QUAPI) and the multilayer multiconfiguration
time-dependent Hartree approach (ML-MCTDH), the changeover between incoherent and
coherent regime is comprehensively investigated \cite{Kast,WangNew,Zhao3}.
The initial set up for the time evolution is that the
spin is in the up state and the bath is a displaced-oscillator state.
Under this initial condition, it was found that, when $s<0.5$
and at zero temperature, the
coherence survives sufficiently strong coupling, i.e., the
oscillatory pattern shown in the dynamics can not be quenched by the
bosonic bath at all. This conclusion is striking but helpful in
understanding the robustness of quantum coherence. However, given
difficulties in achieving an accurate theoretical treatment of the
sub-ohmic bath, much remains to be explored on the time and
temperature dependences of the quantum coherence.
Due to the relative ease of its theoretical treatments, the Ohmic bath has been the focus of
many research efforts, and various numerical methods have been employed, such as the
QUAPI method \cite{PRB94,PRA2}, the real-time
renormalization group \cite{PRB3}, the ML-MCTDH approach\cite{JCP}, and the non-Markovian
noninteracting blip approximation \cite{SBM2,JPCL}. Extensive
investigations of the SBM have also been carried out on low-temperature
dynamics, the crossover from nonadiabatic to adiabatic
behavior \cite{PRB94}, the effect of a non-Markovian environment in the weak
coupling regime \cite{JPCL,NJP}, and entanglement dynamics under the
dephasing \cite{PRA}. The few reports
on the sub-Ohmic bath have typically considered the SBM in thermal equilibrium \cite{ChinPRB,PRB2}.
We note that most QUAPI results \cite{PRB2} fall in the Fermi ``golden rule" regime,
which corresponds to the sub-Ohmic model with relatively strong coupling,
rendering them a useful benchmark for checking validities of other approximate methods.

If the bosonic bath is treated semiclassically,
numerous mixed quantum-classical algorithms are
applicable to SBM \cite{CP,CP2,PRE,JCP2,JCP3,JCP4,JCP_SBM1,JCP_SBM2}.
Indeed, since the bosonic bath occupies an infinite Hilbert space,
the semiclassical treatments are more
efficient and straightforward. The celebrated surface hopping
algorithm \cite{Tully1}, which, thanks to its high efficiency,
has seen wide applications in the field of
chemical physics \cite{Tully2},
is frequently invoked to treat the
SBM \cite{JCP_SBM1,JCP_SBM2}. However, the traditional algorithm of
surface hopping, when applied to the SBM, is rather inadequate as the
quantum coherence needs to be considered additionally \cite{JCP_SBM2}.
By employing the Davydov D$_1$ ansatz,\cite{D1} we develop in this work a novel surface
hopping algorithm for the SBM.
As a semi-classical approach for
studying energy transport in deformable molecular chains, the hierarchy of
Davydov ans\"{a}tze were put forth by Davydov and others as three trial
wave functions, which are known as the D$_1$, $\tilde{\rm D}$, and D$_2$
trial states \cite{Zhao1,Zhao2}. The most sophisticated of the three,
the D$_1$ ansatz is quite accurate and
possesses a compact form suitable to be enlisted in a surface hopping
algorithm, in which the influence of temperature on quantum coherence can be taken
into account naturally.

This paper is organized as follows. The surface hopping algorithm
and the Davydov D$_1$ ansatz for the SBM are introduced in
Section II. Simulation results using the surface hopping algorithm are presented in Section III,
where a comparison of various approaches is given together with discussions
of calculated hopping rates. Conclusions are drawn in Section IV.

\section{Methodology}

\subsection{Davydov D$_1$ ansatz}

We first give a brief introduction to the the Davydov D$_1$ ansatz,
which had been previously applied to the Holstein molecular crystal
model \cite{Zhao1,Zhao2}. Very recently, the ansatz has been successful used to tackle
the sub-Ohmic SBM, revealing an excellent precision of the
trial wave function in the strong coupling
regime \cite{Zhao3}. The D$_1$ ansatz takes the form of a linear
superposition of coherent states as
\begin{eqnarray}
|D_1(t)\rangle=&&\sum_nA_n(t)|n\rangle\nonumber\\&&\otimes\exp\sum_l(B_{n,l}(t)b^{\dag}_l-B^{*}_{n,l}(t)b_l)|0\rangle_b,\label{trial}
\end{eqnarray}
where $A_n(t)$ are the spin variational parameters denoting the occupation
probability of the spin state, $B_{n,l}(t)$ are the corresponding
bosonic displacements for the $l$-th mode,
$n$ takes two values, $+$ and $-$, to
denote the up and down spin states, respectively, and $|0\rangle_b$
is the bosonic vacuum state.
A logarithmic discretization approach is adopted for the bosonic modes, and 500 modes are taken into account.
Equations of motion can be derived for
$A_n(t)$ and $B_{n,l}(t)$, and a key step of the Dirac-Frenkel time-dependent variation is
the projection of the deviation vector $|\delta(t)\rangle$, defined as
\begin{eqnarray}
|\delta(t)\rangle \equiv (i\frac{\partial}{\partial t}-\hat{H})|D_1(t)\rangle ,
\end{eqnarray}
onto the states $|n\rangle\otimes U_n^{\dag}|0\rangle_b$ and
$|n\rangle\otimes U_n^{\dag} b_l^{\dag}|0\rangle_b$, where
$U_n^{\dag}\equiv\sum_l(B_{n,l}(t)b^{\dag}_l-B^{*}_{n,l}(t)b_l)$. This
is equivalent to a minimization procedure that keeps the magnitude of
the deviation vector $|\delta(t)\rangle$ at a minimum at all
times. The equation of motion for $A_\pm(t)$ can then be expressed
as
\begin{eqnarray}
-i\frac{\partial}{\partial
t}A_\pm(t)&=&A_\pm (t)\sum_l[\frac{i}{2}(B^*_{\pm,l}(t)\frac{\partial}{\partial
t}B_{\pm,l}(t)-c.c.)\nonumber\\&-&\omega_l|B_{\pm,l}(t)|^2\mp\frac{\lambda_l}{2}(B_{\pm,l}(t)+c.c.)]\nonumber
\\&\mp&\epsilon/2A_{\pm}(t)+\frac{\Delta}{2}A_{\mp}(t)S_{\pm,\mp},\label{eomA}
\end{eqnarray}
where $S_{n,n'}\equiv\langle 0|U_n U_{n'}^{\dag}|0\rangle$. For
$B_{\pm,l}(t)$, the equation of motion reads,
\begin{eqnarray}
-iA_\pm(t)\frac{\partial}{\partial
t}B_{\pm,l}(t)&=&-\omega_lA_\pm(t)B_{\pm,l}(t)\mp\frac{\lambda_l}{2}A_\pm(t)\nonumber\\&+&\frac{\Delta}{2}A_{\mp}(t)S_{\pm,\mp}(B_{\mp,l}-B_{\pm,l}).\label{eomB}
\end{eqnarray}
At zero temperature, this procedure alone yields the time evolution
of the spin-boson system given a certain initial condition.
However, as we are interested in temperature effects, it is
necessary to consider the surface-hopping algorithms. As the SBM
is a two-level system coupled with a bath of bosons, there are only
two surfaces available, each with a large number of bosonic
trajectories. The integration of the surface hopping
algorithm with the D$_1$ ansatz is elaborated in the next
subsection.

\subsection{Surface hopping algorithm}

In the traditional surface-hopping algorithms, the electron is
allowed to hop between two adiabatic surfaces, each with a set of
independent bosonic trajectories. Based essentially on mixed quantum-classical approaches at a mean-field level,
influence of phonons on the quantum coherence can not be properly described in those algorithms \cite{Tully2}.
On the other hand, as described above, the
Davydov D$_1$ ansatz is a localized variational wave function, in which the
bosonic trajectories are naturally positioned on the diabatic surfaces, and can be calculated in a quantum-mechanical manner. To tackle these concerns,
we devise a procedure for the calculation of hopping between the
diabatic surfaces which will be described as follows.

We first consider the initial condition for the time evolution and choose
the one with the spin in the up state. The randomly-chosen displacements
$B_{n,l}(0)$ are uniformly
distributed within $[0,d]$, where the upper bound $d$ depends on the
temperature $T$, and can be determined by equating the potential
energy to $k_{\rm B}T$ with $k_{\rm B}$ the Boltzmann constant, i.e.,
$\sum_{n,l}\lambda_lB_{n,l}(0)\sim k_{\rm B}T$. This assumption is
equivalent to setting the bosonic displacement dynamically
disordered \cite{TroisiDynamic}. Within this initialization, we
prepare a number of distinct
trajectories (in this work, about 100 trajectories are taken).
The final results are then obtained after performing an
ensemble average thereby accounting for the
temperature effect in our theory.

The equations of motion for
$A_n(t)$ and $B_{n,l}(t)$, Eqs.~(\ref{eomA}) and (\ref{eomB}), can be subsequently
utilized to simulate the time evolution of the system. As an
essential step of the surface hopping algorithm, after each
discrete time interval $\Delta t$ (in this work, $\Delta
t=0.1$ is taken), a spin flipping is allowed in a probabilistic manner.
The switching criterion consists of 1) generating a random number
$\xi$ uniformly distributed in $[0,1)$, and 2) deciding a surface hop is to take place if
\begin{equation}
\xi<\left\{
\begin{array}{ll}
\rho_-(t)\exp[-(\epsilon_--\epsilon_+)/k_{\rm B}T], & \epsilon_->\epsilon_+,  \\
\rho_-(t), & \epsilon_-\leq\epsilon_+,
\end{array}\right.
\end{equation}
where $\rho_-(t)$ is the probability the spin resides in the down
state, and $\epsilon_+$ and $\epsilon_-$ are the total energies for
the spin in up and down state, respectively. With every instance of
spin-flipping, the notation is inverted. This criterion is
based on the Miller-Abrahams formula in which the back and forward
rates obey the detailed balance \cite{MA}. If this criterion is
satisfied, a spin surface hop takes place, otherwise the system continues to evolve until a surface
hop occurs. After the surface hopping, the spin state will quickly relax
to the lowest energy point on the surface while keeping relevant phase information.
The traditional treatment of surface-hopping also
considers an adjustment to the bosonic displacement and
velocity \cite{Tully1,comment1}. Here, since the bosonic part evolves
fully quantum mechanically between the hops, it is not necessary to follow such
adjustments. If the energy change during the surface hopping process
is too large compared to the temperature, a damping term may be
added to Eq.~(\ref{eomB}). However, within the current model and
parameters, we have ascertained (with results shown elsewhere \cite{unpub}) that
such an adjustment is not essential.

\subsection{Novelties and properties of our algorithm}

We list below the novelties and properties of our proposed surface-hopping algorithm.
It is our hope that suitable extensions of this
algorithm can be found in other related models.

(1) The proposed method is capable to handle properly the phase information of the spin during time propagation.
Typically, in the mixed quantum-classical algorithms, it is
cumbersome to take into
account quantum coherences appropriately as the classical dynamics of the bosons
and the quantum dynamics of the spin appear fundamentally incompatible \cite{Tully2}.
In our algorithm, since both the spin and
boson parts are treated quantum mechanically, such problems are
completely circumvented. Similarly, in using the master equations,
the memory effect is also an important issue to
consider \cite{JPCL,NJP}. In our approach, the memory effect
is included in the trajectories.

(2) All the trajectories on the surfaces are taken into
consideration in the theoretical framework, and their individual
influences on the spin motion are accounted for. In some approaches,
the bosonic bath is assumed to be in thermal equilibrium during time
evolution \cite{ChinPRB,PRB2}, and the eventual tunneling between
different trajectories, which essentially gives rise to the quantum
effect, may be improperly quenched. In our algorithm, however,
the fluctuations among trajectories are included for a proper
simulation of the temperature effect.

(3) Our algorithm does not strictly obey the detailed balance just as the traditional surface hopping algorithms without any artificial modification \cite{detailB}. We use the Miller-Abrahams formula, which obeys the Boltzmann distribution and is frequently adopted for incoherent hopping \cite{Bassler}, and choose to refrain from any interventions. This is quite different from other surface hopping algorithms that artificially adjust velocities in each hopping event due to the requirement of energy conservation. Despite its importance and desirability, energy conservation in a surface hopping event, due to various technical reasons, may not be able to be satisfied, and in some cases, may not even be absolutely necessary in a surface-hopping algorithm. For instance, there are algorithms in which velocity adjustment is taken only in hopping events that are classically allowed, or no such adjustment is made at all \cite{adjust}, because of difficulties in simultaneously satisfying the energy conservation requirement and the self-consistency condition of the mixed quantum-classical propagation. Inconsistencies can occur when there is insufficient kinetic energy of the nuclei to compensate at an hop from a lower electronic state to a higher one, and a consensus remains elusive on how to get around such problems. Here, as we are dealing with a quantum bath of bosons, artificial adjustment of velocities is no longer straightforward, although tampering of the boson displacements for each mode is among our future options. As the first step in our algorithm development, the Miller-Abrahams formula is employed, and within this optimization, temperature divergence is avoided
as opposed to other surface hopping algorithms \cite{Tully2}.

(4) The roles of the bosons are twofold: to initiate the thermally assisted
hopping process, and to act as scattering centers for the wave packet
propagation in a bandlike process \cite{yaotd}.
As opposed to usual schemes for including
the temperature effects which often incorporate only one of the two roles,
in our algorithm we take both into consideration. To be more
specific, the former role originates in thermally influenced
hopping between two surfaces, while the latter arises from the initial
distribution of trajectories. The two dominate in different
regimes, a fact that turns out to be advantageous for the current
algorithm.

(5) Traditional surface hopping algorithms handle the bosons classically due to difficulties associated with a quantum-mechanical treatment. The Davydov D$_1$ ansatz reduces the Schr\"{o}dinger equation in an infinite-dimension bosonic Hilbert space into a manageable set of equations of motion, therefore making it possible to embed the quantum evolution of the spin-boson system in a surface hopping algorithm.
The calculations are carried out on a Linux-based computing cluster, with each CPU being a 2.67GHz Intel processor.
In the test runs, one CPU hour is sufficient to achieve a convergent result with more than 1000 samplings for a given set of control parameters. This high computational efficiency makes it feasible to extend our algorithm to large, realistic systems.

(6) The algorithm presented here can be applied to other relevant
models in a straightforward manner, e.g., to the Holstein molecular
crystal model on a lattice of a finite number of sites. The exciton
in the Holstein Hamiltonian can be initially created on one site and
then propagate on the lattice in a quantum mechanical manner, and after
a certain amount of time, surface hopping will take place.
In devising a criterion for this incoherent
surface hopping and searching for a target site the exciton will hop
to, one needs to consider the sum of the probabilities on all other
sites and compare it to a random number within [0,1] \cite{yaotd}.
The whole procedure will then be similar to the current one.

\section{Results and discussions}

\subsection{Comparison with numerically exact results}

\begin{figure}
\includegraphics[angle=0,scale=0.9]{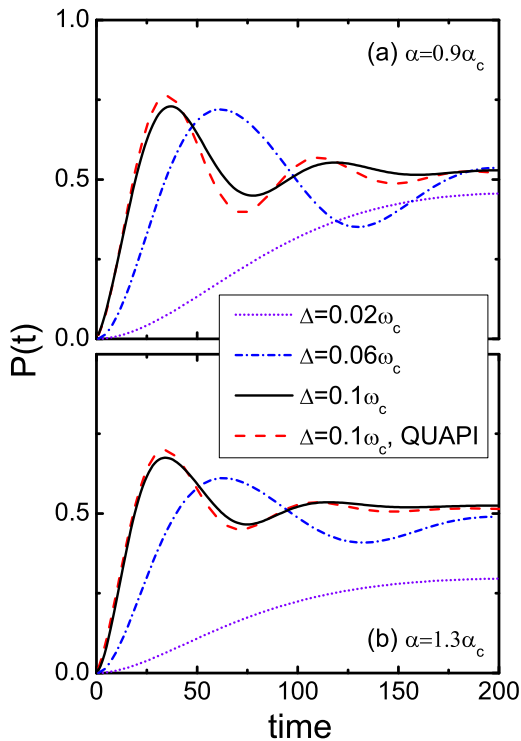}
\caption{Evolution of the down spin population for (a) $\alpha=0.09\alpha_c$ and (b) $\alpha=0.13\alpha_c$ and $k_{\rm B} T=0.1\Delta$.
The QUAPI results are extracted from Ref. \cite{PRB2}.}\label{QUAPI}
\end{figure}

In order to assess the reliability of our method, results obtained here are compared in Fig.~\ref{QUAPI} with those from the numerically exact QUAPI method, which are extracted from Ref.~\cite{PRB2}. Here $\alpha_c~(\sim0.022)$ is the critical coupling strength for the delocalized-localized transition, and two values of $\alpha$, i.e, $0.9\alpha_c$ and $1.3 \alpha_c$, are taken for comparison. It is found that the QUAPI results and ours are in good agreement at short times, but a slight deviation between the two emerges at long times. In addition, visibly better agreements are achieved in the strong coupling regime due to the coherent-state structure of the D$_1$ ansatz. The comparisons here are made at a relatively low temperature, and as is the case for most surface hopping algorithms, we expect that our method becomes more reliable at higher temperatures.

\begin{figure} [b]
\includegraphics[angle=0,scale=0.8]{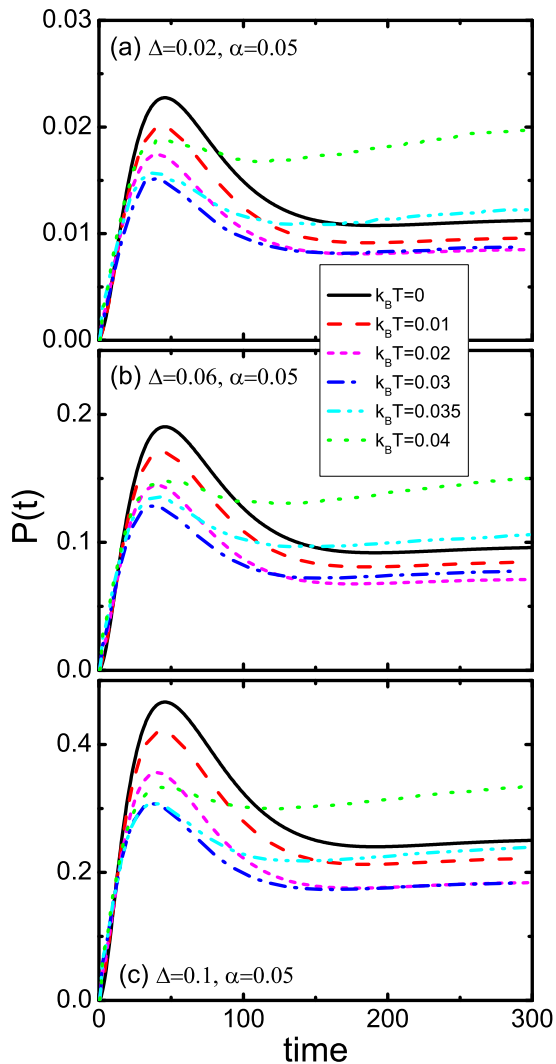}
\caption{Evolution of the down spin population for $\alpha=0.05$, six
temperatures and three values of the tunneling constant without bias.}\label{unbias1}
\end{figure}

\subsection{Evolution of population}

\begin{figure}
\includegraphics[angle=0,scale=0.8]{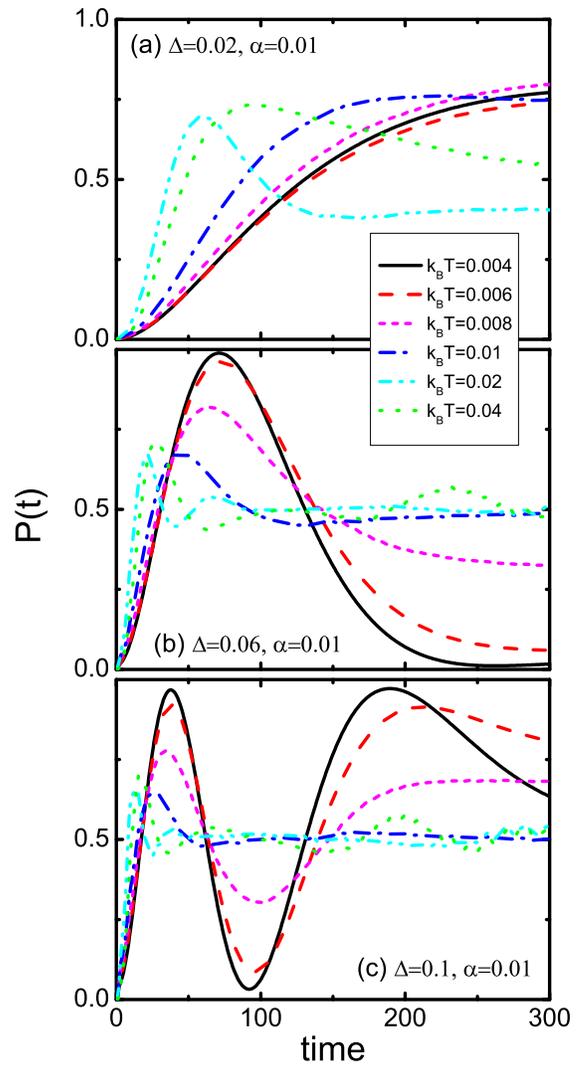}
\caption{Evolution of the down spin population for $\alpha=0.01$, six
temperatures and three values of the tunneling constant without bias.}\label{unbias2}
\end{figure}

We shall investigate the sub-Ohmic case of $s=1/4$, as a first example for
studying the population evolution. Figs.~\ref{unbias1} and
\ref{unbias2} display the results for cases of strong and weak spin-boson coupling,
respectively. From comparing the population evolution in Fig.~\ref{unbias1},
it is clear that the low-temperature behavior shows substantial deviations from its high-temperature
counterpart around $k_{\rm B}T=0.03$ for the
strong coupling case.
At low temperatures, mainly
damped oscillations are observed, and with an increase in
$\Delta$, a population maximum grows significantly. More
importantly, the effect of the temperature hinders population propagation,
implying that the temperature acts as the scattering source for
the spin flipping process. At high temperatures, the down-spin
population does not enter a steady state, but continues to grow
at long times, a fact that has its origin in the
thermally assisted hopping.
Thanks to thermal agitations, the spin
flipping survives the strong damping of the bosonic bath at high temperatures. It is
thus clear that our surface hopping algorithm can
simultaneously include negative and positive temperature
dependencies. This points to possible important applications of our algorithm to complicated
transport problems in molecular systems, where both the coherent and
incoherent mechanisms coexist \cite{yaotd}.

\begin{figure}
\includegraphics[angle=0,scale=0.7]{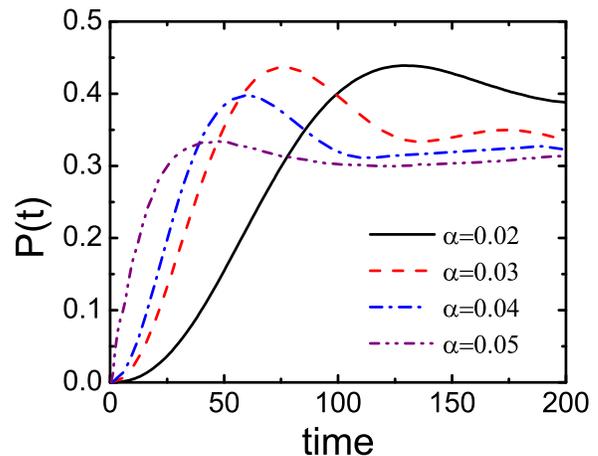}
\caption{Evolution of the down spin population for four values of spin-boson coupling. The parameters are $\Delta=0.1, k_{\rm B} T=0.04$.}\label{alphaChange}
\end{figure}

\begin{figure}
\includegraphics[angle=0,scale=0.7]{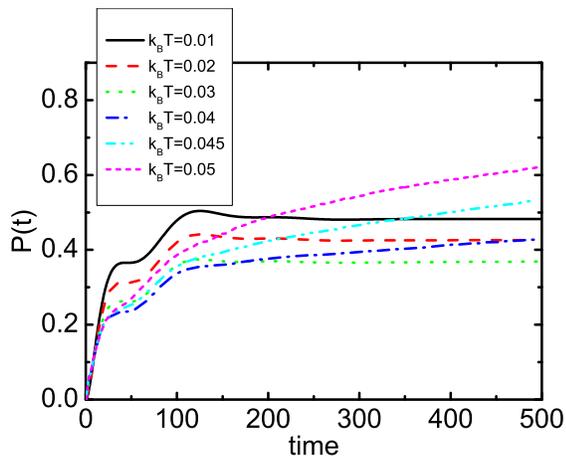}
\caption{Evolution of the down spin population for six
temperatures under a bias of $\epsilon=0.04$.
Other control parameters are $\Delta=0.1, \alpha=0.01$.}\label{bias}
\end{figure}

For the case of weak coupling, the difference of two temperature regimes
disappears, since the influence of the bath becomes weaker. With an
increase in temperature, the amplitude of the oscillation decreases as shown in Fig. \ref{unbias2}(c), and the population reaches $0.5$ at long
times. This result implies that quantum coherence could be largely quenched at elevated temperatures.
In order to show this more clearly, we display in Fig.~\ref{alphaChange} the dependence of the
down-spin population evolution on $\alpha $ for $k_{\rm B}T=0.04$. It can be seen that the oscillatory behavior of the population is suppressed as
$\alpha$ is increased,
which is quite different from the zero-temperature case {\cite{Zhao3}}.
In Fig.~\ref{bias}, we also show the
population evolution of the down-spin state with bias $\epsilon=0.04$.
The overall trend is
found to be similar to that in Fig.~\ref{unbias1}, i.e.,
two temperature regimes exist even under bias \cite{comment2}.
In particular, the line shape in the presence of the bias exhibits a hint of exponential
decay, similar to that in the incoherent hopping regime \cite{JCP_SBM2}, an analogy that may
prove useful in extracting
incoherent hopping rates from our results.

\subsection{Comparison with Marcus formula}

The spin dynamics is nearly coherent before the onset of surface hopping,
and the calculated $\rho_-(t)$ could be regarded
as the coherent hopping probability for the spin, depending only on
the tunneling frequency and the spin-bath coupling strength. However, at high temperatures,
hopping processes in molecular systems are
always incoherent, and the hopping rate strongly depends on the
disorder in molecular structure and configuration. Usually, one can
use a rate equation
\begin{eqnarray}
\frac{{\rm d}P(t)}{{\rm d}t}=-kP(t)\label{rateEq}
\end{eqnarray}
to estimate the incoherent hopping rate, in which $P(t)$ is the
diabatic population of the final state and $k$ is the hopping
rate \cite{JCP_SBM2}. However, the problem can be considerably more complicated, since
$k$ may be time dependent due to the coherent motion. Therefore,
we follow the general approach of using the Fermi ``golden rule" to
calculate the incoherent hopping rate \cite{Fermi}, i.e.,
\begin{eqnarray}
k=2\pi\sum_f\sum_i\frac{{\rm e}^{-\epsilon_i/k_{\rm B}T}}{Z}|\langle
f|\hat{V}|i\rangle|^2\delta(\epsilon_i-\epsilon_f),
\end{eqnarray}
with $i$ and $f$ denoting the initial and final state for the
hopping process and $Z$ being the partition function. In our case,
especially, the initial state for the spin is taken to be
$|+\rangle$, while the final spin state is $|-\rangle$. To make up
the energy change during spin flipping, the corresponding bosonic
states will change adiabatically, so that the operator $\hat{V}$
includes both spin flipping term $\sigma^x$ and the creation and
annihilation operators for the bosons. For the spin part only, the term
$Z^{-1}\sum_i{\rm e}^{-\epsilon_i/k_{\rm B}T}|\langle f|\hat{V}|i\rangle|^2$
can be expressed by the trace over the initial states, i.e., ${\rm
Tr}_i(\sigma^x\rho_f\sigma^x)$ with $\rho_f\equiv\left(
\begin{array}{ll}
0,~~0\\0,\rho_-(t)
\end{array}\right)$, which is proportional to $\rho_-(t)$. For the other parts, we
introduce a so-called attempt-to-escape frequency $\nu_0$ to include
all the \textit{extrinsic} factors and the bosonic motions that are
much slower than the spin flipping. As we are interested in the long
time average, $\nu_0$ is set to 0.1. Hence, $k$ in Eq.~(\ref{rateEq})
becomes equal to $\nu_0\rho_-(t)$, and by integrating over Eq.~(\ref{rateEq})
we can derive the average waiting time as
\begin{eqnarray}
\langle
t_w\rangle=\int_0^{\infty}\exp(-\nu_0\int_0^t\rho_-(\tau){\rm
d}\tau){\rm d}t.
\end{eqnarray}
The average incoherent hopping rate can then be expressed as
$k=1/\langle t_w\rangle$.

\begin{figure}
\includegraphics[angle=0,scale=0.7]{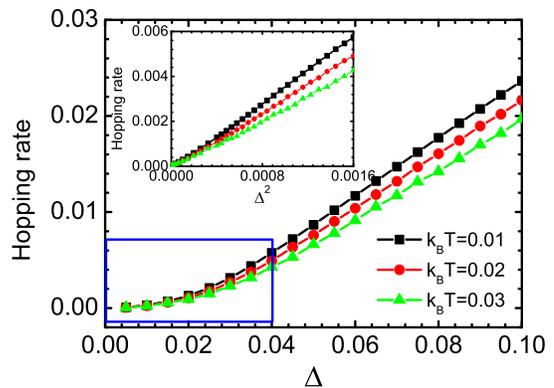}
\caption{The hopping rate versus $\Delta$ and $\Delta^2$ (inset) for
$\alpha=0.01$ for three temperature with bias $\epsilon=0.04$.
The blue box denotes the regime that the hopping rate is
proportional to $\Delta^2$.}\label{rate}
\end{figure}

In the traditional surface hopping algorithm, such as its
semiclassical versions, the hopping rate is shown to be linearly dependent on
$\Delta$ in the small $\Delta$ regime \cite{JCP_SBM2}. In the
Marcus regime, though, it is well known that the hopping rate follows
a $\Delta^2$ scaling. This deviation is known to occur in the
absence of decoherence during the surface hopping \cite{JCP_SBM2}. In
our algorithm here, however, as both the spin and bosons are
treated fully quantum-mechanically between the hops, we expect our results to capture
the proper scalings. Shown in Fig.~\ref{rate} is the $\Delta$
dependence of the hopping rate with a bias of $0.04$ and at high
temperatures, which are the typically applicable regimes of the Marcus
theory. It is found that, when $\Delta$ is larger than $0.04$, i.e.,
out of the perturbative regime, the hopping rate is linearly dependent on $\Delta$,
similar with the traditional case. However,
when $\Delta$ is smaller than $0.04$, the results clearly demonstrate the
$\Delta^2$ scaling, which is perfectly consistent with the Marcus
formula. Therefore, we can conclude that the present treatment
reliably captures the essential behavior of the incoherent hopping
processes.

\subsection{From sub-Ohmic to Ohmic case}

\begin{figure}
\includegraphics[angle=0,scale=0.6]{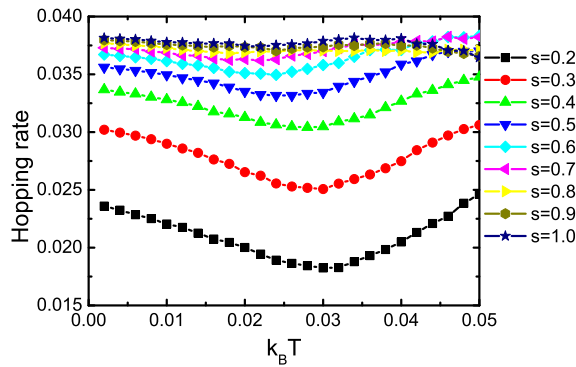}
\caption{The hopping rate versus temperature for various $s$ with
$\Delta=0.1, \alpha=0.05$.}\label{sDepen}
\end{figure}

Phenomenologically, negative temperature dependencies are
referred to as a feature in coherent transport, while positive temperature dependencies
are associated with the incoherent hopping regime. Earlier reports indicate that
$s=0.5$ is a transition point that separates coherent and incoherent regimes
when the coupling is sufficiently strong, indicating an interesting $s$ dependence in this model \cite{Kast}.
In this
work, we can further examine the transition in the presence of
temperature. Fig.~\ref{sDepen} depicts the hopping rate as a function of the
temperature for a number of $s$ values from $0.2$ to $1.0$. In
the low-temperature regime (e.g., $ k_{\rm B} T <0.025$), a coherent transport regime can be identified
for $s<0.7$, however for larger values of $s$, the insensitivity of the
hopping rate dependence on temperature prevents a clear
interpretation, implying that thermal fluctuations do not
influence the spin dynamics as the system becomes
more incoherent and localized. This result is quite
interesting, and it can be studied in a more comprehensive manner in other
related models. For example, if we are dealing with the lattice
model, such as the Holstein molecular crystal model, such an effect may give rise to
new physics.

\section{conclusions and discussion}

In summary, we have developed a surface hopping algorithm employing the fast, accurate
Davydov D$_1$ ansatz originally devised to simulate polaron dynamics in the Holstein molecular
crystal model.
The primary advantage this method
offers is that
both the coherent and incoherent hopping processes are taken into
consideration as the procedure is carried out in a
mixed quantum-classical manner.
We have applied this method to the sub-Ohmic
spin-boson model to calculate the population evolution, and thus the
hopping rate between spin up and down states. In the strong coupling regime, a transition
from coherent to incoherent transport is found to take place. Results obtained in this work are compared
with those from the Marcus theory, and a
satisfactory agreement is reached; in particular, the $\Delta^2$ scaling of the hopping rate in the Marcus regime is recovered.
Lastly, the dependence of the hopping rate on $s$ has been studied in detail.

The surface-hopping algorithm has been under development for more than two decades, but many issues remain unsettled. Realizations of the algorithm proposed by Tully vary in choice and degree of sophistication of the hopping criteria. A proper treatment of quantum coherence, for example, is recently found to be essential in reproducing the Marcus golden-rule rate. Modifications of the algorithm, such as the sudden quenching of electronic coherence and adjustment of nuclear velocities, have been made to tackle these issues.
In this work, an alternative means to circumvent the problem is proposed by embedding the Dirac-Frenkel time-dependent variation in the surface hopping algorithm.
Despite the approximate nature of the variational theory, quantum coherence is expected to be properly taken into account in such an approach.
The sub-Ohmic bath is chosen first as it is characterized by strong system-bath coupling, and is known to be accurately described by the Davydov trial states \cite{Zhao3}. Comparisons of our results with those from the numerically exact QUAPI method for the unbiased case confirms the robustness of our approach in this ``golden rule" regime \cite{SBM1}.
To extend applications of our algorithm to additional, perhaps realistic systems, care needs to be taken on choosing the hopping criteria and adopting possible adjustments to the variational parameters at each hop. Validity of the algorithm needs to be assessed on a case by case base, similar to various developments of the traditional surface hopping algorithms.

\begin{acknowledgments}
This work was supported by the Singapore National Research
Foundation through the Competitive Research Programme (CRP) under
Project No.~NRF-CRP5-2009-04. One of us (YY) is also supported in
part by the National Natural Science Foundation of China and the National
Basic Research Program of China (2009CB929204 and 2012CB921401).
The authors thank an anonymous reviewer, Sebastian Fernandez Alberti, and Vladimir Chernyak for useful discussion.
\end{acknowledgments}


\begin{thebibliography}{99}
\bibitem{comment1} The requirement of velocity adjustments in a surface hopping event is a subject that has undergone extensive discussion, and it has not been completely demonstrated that the energy conservation is essential at each hop \cite{adjust}. Thus, the issue of energy conversation in a hopping event should be carefully examined case by case.
\bibitem{comment2}The sub-Ohmic model, which is characterized by strong system-bath coupling, falls within the Fermi ``golden rule" regime, where most QUAPI results are obtained. We expect our biased results to be as good as the unbiased ones despite the absence of a direct comparison with corresponding biased QUAPI results.
\end{thebibliography}
\end{document}